\documentclass[aps,prb,reprint,preprintnumbers,showpacs,showkeys,superscriptaddress]{revtex4-2}
\usepackage{amsmath,amssymb,bm,mathrsfs,braket,color,blkarray,ulem}
\usepackage[pdftex]{graphicx}
\usepackage{srcltx}
\usepackage[colorlinks=true,citecolor=blue,linkcolor=blue]{hyperref}

\begin{document}
\title{Theory of spin magnetization driven by chiral phonons}

\author{Dapeng Yao}
\email{yao.dapeng@stat.phys.titech.ac.jp}
\affiliation{Department of Physics, Institute of Science Tokyo, 2-12-1 Ookayama,
Meguro-ku, Tokyo 152-8551, Japan}
\altaffiliation[Former address: ]{Department of Physics, Tokyo Institute of Technology, 2-12-1 Ookayama, Meguro-ku, Tokyo 152-8551, Japan}

\author{Shuichi Murakami}
\email{murakami@stat.phys.titech.ac.jp}
\affiliation{Department of Physics, Institute of Science Tokyo, 2-12-1 Ookayama, Meguro-ku, Tokyo 152-8551, Japan}
\altaffiliation[Former address: ]{Department of Physics, Tokyo Institute of Technology, 2-12-1 Ookayama, Meguro-ku, Tokyo 152-8551, Japan}
\affiliation{International Institute for Sustainability with Knotted Chiral Meta Matter (WPI-SKCM$^\text{2}$),
Hiroshima University, 1-3-1 Kagamiyama, Higashi-Hiroshima, Hiroshima 739-8526, Japan
}

\begin{abstract}
We construct a general theory of spin magnetization driven by chiral phonons under an adiabatic process, in which atoms rotate around their equilibrium positions with a low phonon frequency. Here the spin magnetization originates from the modulated electronic states with spin-orbital coupling by atomic rotations.
Under the adiabatic approximation, the time-dependent spin magnetization can be calculated by a Berry-phase method. In this paper, we focus on its time average, which is evaluated by assuming that the phonon displacement is small. As a result, the time average of the spin magnetization is concisely formulated in the form of the Berry curvature defined in the phonon-displacement space as an intrinsic property of atomic rotations.
Our formula for spin magnetization reflects the chiral nature of phonons, and is convenient for $ab$ $initio$ calculations.

\end{abstract}
\maketitle

\section{Introduction}
\label{secI}
Chiral phonons accompanying atomic rotations represent circularly polarized modes, which carries phonon angular momentum~\cite{Zhang2014,Zhang2015,Chen2018,Chen2019,Wang2022,Zhang2022,Zhang2023,Tsunetsugu2023}.
Such phonons have been confirmed by recent experiments~\cite{Zhu2018,ChenXT2019,Grissonnanche2020,Tauchert2022,Ueda2023,Ishito2023}, and their various aspects have been revealed, such as the large phonon magnetic moment owing to both electron spin and orbital contributions~\cite{Juraschek2019,Cheng2020,Hamada2020,Xiao2021,Ren2021,Geilhufe2021,Baydin2022,Xiong2022,Juraschek2022,Luo2023,Hernandez2023,Chaudhary2024}, the induced electric current~\cite{Yao2022} and spin current~\cite{Fransson2019,Kim2023,Li2024,Ohe2024,YaoAPL2024,Funato2024}, the conversion between chiral phonons and magnons~\cite{Thingstad2019,Yao2024}, the light-driven phonon chirality in a paramagnet~\cite{Ren2024} and a ferromagnet~\cite{Kahana2023}, and the phononic switching of magnetization~\cite{Davies2023}.
These works indicate the angular momentum transfer between phonons and electrons.

If the phonon magnetic moment comes only from the rotating motion of the nuclear charge, it does not match the realistic value of the anomalously large phonon magnetic moment measured in the recent experiments~\cite{Cheng2020,Baydin2022}.
Instead, the atomic rotation modulates the electronic states, which induces electronic magnetic moment. It can be calculated on the assumption that this modulation is an adiabatic process, by using the method in Ref.~\cite{Berry1984}. It is shown that the atomic rotation induces spin magnetization~\cite{Hamada2020} and orbital magnetization~\cite{Xiao2021,Ren2021,Trifunovic2019} due to geometric effect.
The atomic rotation also induces the electric current~\cite{Yao2022}, and the magnetic spin excitation~\cite{Yao2024} as a geometric effect.

In this paper, by means of the time-dependent description of the Berry-phase method applied to chiral phonons~\cite{Hamada2020}, we first derive a general formula for the time average of the spin magnetization, which is proportional to the phonon angular momentum. 
It is concisely expressed in terms of the Berry curvature defined in the phonon-displacement space.
We then introduce a tight-binding (TB) model of a two-dimensional (2D) honeycomb lattice with the Rashba spin-orbital coupling (SOC) to verify our formalism. Here we consider the chiral phonon at the $\Gamma$ point, which dynamically modulates the electronic Hamiltonian.
We show that the results from our formalism well agree with those by the time-dependent description~\cite{Hamada2020} when the phonon displacement is small.
We next discuss that how our formalism is applicable for general cases, in which the degrees of freedom of phonon displacement contains more than one single atom.
Our concise formula for spin magnetization reflects the chiral nature of phonons in rotational modes while it vanishes in simple vibration modes.
Since the Berry curvature here can be easily obtained by the calculation of electronic states, our formalism is convenient for $ab$ $initio$ calculations.

The remainder of the paper is organized as follows. We first construct a formalism of spin magnetization driven by chiral phonons in the case of one single atomic rotation within the unit cell in Sec.~\ref{secII}. In Sec.~\ref{secIII}, we calculate the spin magnetization for a TB model of a 2D honeycomb lattice with Rashba SOC to verify our formalism. Section~\ref{secIV} discuss the applicability of our formalism to general cases. We conclude the paper in Sec.~\ref{secV}.
Supplementary details of our calculation are placed in Appendix.

\section{Formalism}
\label{secII}
In this section, we propose a general theory of spin magnetization driven by chiral phonons with an adiabatic approximation. Here we study a band insulator and we assume that the phonon frequency $\omega$ is much lower than the band gap. It means that the electronic system is adiabatically modulated by atomic rotations, leading to a geometric effect on spin magnetization~\cite{Trifunovic2019,Hamada2020}.

For simplicity, we consider the case where the unit cell contains a single atom, and the phonon displacement of this atom is within the $xy$ plane, characterized by $\bm u=(u_x,u_y)$ with $u_{\delta}$ $(\delta=x,y)$ periodically changing with time. Then $u_{\delta}$ is regarded as a parameter in the Bloch Hamiltonian $\mathcal{H}(\bm k,\bm u)$, which describes the dynamics of electrons affected by atomic rotations.
We assume that the modulation due to atomic rotations does not close the electronic band gap.
Within the adiabatic approximation, the formula for the time dependence of the spin magnetization is derived in Ref.~\cite{Hamada2020}, where the geometric phase due to the dynamic modulation contributes.
In this paper, we focus on its time average. By taking a time average of the spin magnetization calculated in Ref.~\cite{Hamada2020}, the geometric spin magnetization for the $\alpha$ component $(\alpha=x,y,z)$ from the $n$-th electronic band is given by
\begin{align}\label{mu_sn}
\mu^{\alpha}_{s,n}=\sum_{\delta}\dot{u}_{\delta}\int_{\text{BZ}}\frac{d\bm k}{(2\pi)^d}\Omega^{(n)}_{B_{\alpha}u_{\delta}}\Big|_{\bm B=0},
\end{align}
where $B_{\alpha}$ is the $\alpha$ component of the Zeeman magnetic field, $\Omega^{(n)}_{B_{\alpha}u_{\delta}}=\partial_{B_{\alpha}}A^{(n)}_{u_{\delta}}-\partial_{u_{\delta}}A^{(n)}_{B_{\alpha}}$ is the Berry curvature for the $n$-th electronic band with $A^{(n)}_{\rho}=i\braket{\psi_n|\partial_{\rho}\psi_n}$ being the Berry connection~\cite{Xiao2009}. The derivation of Eq.~(\ref{mu_sn}) is placed in Appendix~\ref{Appe_A}.
Since the displacement $u_{\delta}$ is typically much smaller compared with the lattice constant, we can expand $\mu^{\alpha}_{s,n}$ in Eq.~(\ref{mu_sn}) near $\bm u=0$ to the first order in $\bm u$, which reads
\begin{align}\label{expansion}
\mu^{\alpha}_{s,n}\approx&\sum_{\delta}\dot{u}_{\delta}\int\frac{d\bm k}{(2\pi)^d}\Omega^{(n)}_{B_{\alpha}u_{\delta}}\Big|_{\bm u=0,\bm B=0} \nonumber\\
&+\sum_{\delta\gamma}\dot{u}_{\delta}u_{\gamma}\int\frac{d\bm k}{(2\pi)^d}\partial_{u_{\gamma}}\Omega^{(n)}_{B_{\alpha}u_{\delta}}\Big|_{\bm u=0,\bm B=0},
\end{align}
where the Berry curvatures are evaluated at $\bm u=0$.
In the second line in Eq.~(\ref{expansion}), one can separate the following term into symmetric and anti-symmetric parts as
\begin{align}\label{sym_asym}
\sum_{\delta\gamma}\dot{u}_{\delta}&u_{\gamma}\partial_{u_{\gamma}}\Omega^{(n)}_{B_{\alpha}u_{\delta}}\nonumber=u_x\dot{u}_x\partial_{u_x}\Omega^{(n)}_{B_{\alpha}u_{x}}+u_y\dot{u}_y\partial_{u_y}\Omega^{(n)}_{B_{\alpha}u_{y}} \nonumber\\
&+\frac{1}{2}\left(u_x\dot{u}_y+u_y\dot{u}_x\right)\left(\partial_{u_x}\Omega^{(n)}_{B_{\alpha}u_{y}}+\partial_{u_y}\Omega^{(n)}_{B_{\alpha}u_{x}}\right) \nonumber\\
&+\frac{1}{2}\left(u_x\dot{u}_y-u_y\dot{u}_x\right)\left(\partial_{u_x}\Omega^{(n)}_{B_{\alpha}u_{y}}-\partial_{u_y}\Omega^{(n)}_{B_{\alpha}u_{x}}\right).
\end{align}
After taking the time average over one phonon period $T$ with $T=\frac{2\pi}{\omega}$, the first term on the right-hand side of Eq.~(\ref{expansion}) and the symmetric part of Eq.~(\ref{sym_asym}) vanish because of the periodicity of $u_{\delta}$, and only the anti-symmetric part [fourth term of Eq.~(\ref{sym_asym})] contributes to the time-averaged spin magnetization as
\begin{align}\label{mu_z}
\braket{\mu^{\alpha}_{s,n}}=\frac{1}{2}J_z^{\text{ph}}\int\frac{d\bm k}{(2\pi)^d}\partial_{B_{\alpha}}\Omega^{(n)}_{u_{x}u_{y}}\Big|_{\bm u=0,\bm B=0},
\end{align}
where $J_z^{\text{ph}}=\frac{1}{T}\int_0^Tdt(\bm u\times\dot{\bm u})_z$ represents the phonon angular momentum divided by the mass of the atom~\cite{Zhang2014,Zhang2015}. Here we used the identity $\partial_{u_x}\Omega^{(n)}_{B_{\alpha}u_y}-\partial_{u_y}\Omega^{(n)}_{B_{\alpha}u_x}=\partial_{B_{\alpha}}\Omega^{(n)}_{u_xu_y}$, and
\begin{align}\label{BC_n}
\Omega^{(n)}_{u_xu_y}=\partial_{u_x}A^{(n)}_{u_y}-\partial_{u_y}A^{(n)}_{u_x}
\end{align}
denotes the Berry curvature defined in terms of the phonon coordinates $(u_x,u_y)$. We show the calculation of the derivative of the Berry curvature $\partial_{B_{\alpha}}\Omega_{u_xu_y}^{(n)}$ at $\bm B=0$ in Appendix~\ref{Appe_B}.
The resulting formula in Appendix~\ref{Appe_B} no longer contains $\bm B$, and therefore it is convenient for practical calculations.
From Eq.~(\ref{BC_n}), the spin magnetization changes its sign when the direction of atomic rotations is reversed, and it means that the direction of atomic rotations is reflected by electron spins~\cite{Hamada2020,Yao2022,Yao2024}.

\section{Tight-binding model calculation}
\label{secIII}
In this section, we introduce a TB model of a 2D honeycomb lattice with SOC, which is a minimal model modulated by atomic rotations. Our calculation results verify the formalism proposed in Sec.~\ref{secII}. 

\subsection{Tight-binding model}
Our TB model is the Kane-Mele Hamiltonian on the honeycomb lattice~\cite{Kane2005,Hamada2020} but without the next-nearest-neighbor SOC terms. It is given by
\begin{align}\label{H0}
\hat{H}_0=~&m\sum_{i}\xi_i\hat{c}_i^{\dagger}\hat{c}_i+t_0\sum_{\braket{ij}}\hat{c}_i^{\dagger}\hat{c}_j \nonumber\\
&+i\lambda_R\sum_{\braket{ij}}\hat{c}_i^{\dagger}\left(\bm s\times\frac{\bm d_{ij}}{|\bm d_{ij}|}\right)_z\hat{c}_j,
\end{align}
where $\hat{c}_i=(\hat{c}_{i\uparrow},\hat{c}_{i\downarrow})^{\mathsf{T}}$ $[\hat{c}^{\dagger}_i=(\hat{c}^{\dagger}_{i\uparrow},\hat{c}^{\dagger}_{i\downarrow})]$ is the electron annihilation (creation) operator at site $i$. The first term describes a staggered on-site potential $m$ with $\xi_{A(B)}=\pm1$ for the A(B) sublattice, the second term denotes the nearest-neighbor (NN) hopping with the hopping amplitude $t_0$, and the third term stands for the Rashba SOC with the parameter $\lambda_R$. Here $s_{\alpha}$ $(\alpha=x,y,z)$ is the Pauli matrices of electron spin, $\bm d_{ij}$ is the NN bond vector from the $i$-th site to the $j$-th, and $\braket{ij}$ represents the summation over the NN sites within the honeycomb lattice.

\begin{figure}
\begin{center}
\includegraphics[width=\columnwidth]{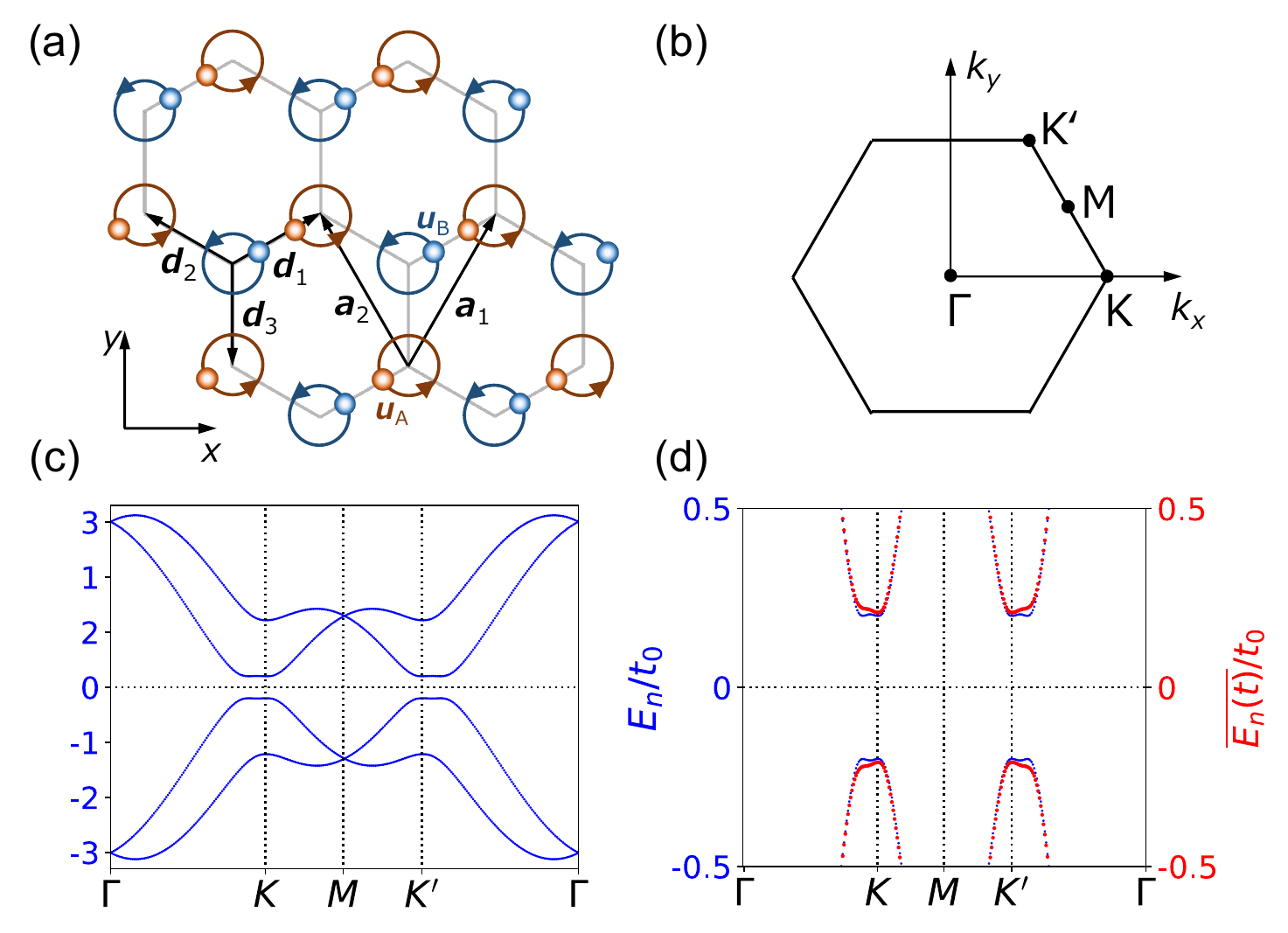}
\caption{(Color online) Calculation results of TB model on 2D honeycomb lattice. (a) Schematic view of a 2D honeycomb lattice with $\Gamma$-optical chiral phonons. Here the sublattices A and B rotate around their equilibrium positions with a phase difference $\pi$, and their displacements are denoted by $\bm u_A$ and $\bm u_B$. The primitive translation vectors are chosen as $\bm a_1=(a/2)(1,\sqrt{3})$ and $\bm a_2=(a/2)(-1,\sqrt{3})$ with lattice constant $a$, and the vectors connecting the NN sublattices are written as $\bm d_1=a_0(\sqrt{3}/2,1/2)$, $\bm d_2=a_0(-\sqrt{3}/2,1/2)$, and $\bm d_3=a_0(0,-1)$ with the NN bond length $a_0=a/\sqrt{3}$. (b) First Brillouin zone with high-symmetry points for the 2D honeycomb lattice. (c) Electronic band structure of the Kane-Mele Hamiltonian in Eq.~(\ref{H0}). (d) Electronic bands are modulated by chiral phonons, leading to a small dynamical energy shift. We compare the band structure without phonons (blue dots) and that with phonons (red dots) after taking the time average. This calculation employs the parameters: $m=0.2t_0$, $\lambda_R=0.4t_0$, and the relative position $u_r$ is 10\% of the bond length $a_0$.}
\label{fig1}
\end{center}
\end{figure}

In the presence of chiral phonons, which make the atoms to rotate around their equilibrium positions, the parameters of the NN terms are modulated because the distance between the NN atoms periodically changes with time due to atomic rotations.
Let $\bm u_i$ and $\bm u_j$ be displacement of the $i$-th and $j$-th atoms which are nearest neighbors, and their relative position changes into $\bm d_{ij}\rightarrow\bm d_{ij}+\bm u$ with $\bm u=\bm u_j-\bm u_i$.
Here we assume that the atomic orbitals at each atoms are isotropic, and thus the hopping amplitudes are determined by the distances between atoms.
Since the distance between the NN atoms changes from $a_0$ to $a_0+\bm u\cdot\bm d_{ij}/a_0$, we assume that the hopping amplitude $t_0$ is modulated by chiral phonons as
\begin{align}
t_0\rightarrow t_0+\delta t_{ij}(\bm u),~\delta t_{ij}(\bm u)=-\frac{t_0}{a_0^2}\bm u\cdot\bm d_{ij},
\end{align}
with the NN bond length $a_0$, leading to the following additional term to the Hamiltonian as~\cite{Hamada2020,Yao2022}
\begin{align}\label{dHt}
\delta\hat{H}_t(\bm u)=\sum_{\braket{ij}}\delta t_{ij}(\bm u)\hat{c}_i^{\dagger}\hat{c}_j.
\end{align}
On the other hand, since $1/|\bm d_{ij}|$ in the Rashba SOC term changes into
\begin{align}
\frac{1}{|\bm d_{ij}|}\rightarrow\frac{1}{|\bm d_{ij}+\bm u|}\approx\frac{1}{a_0}\left(1-\frac{1}{a_0^2}\bm u\cdot\bm d_{ij}\right),
\end{align}
we assume that the Rashba SOC Hamiltonian acquires the following additional terms~\cite{Hamada2020}
\begin{align}\label{dHR}
\delta\hat{H}_R(\bm u)=&~i\frac{\lambda_R}{a_0}\sum_{\braket{ij}}\hat{c}_i^{\dagger}\left(\bm s\times\bm u\right)_z\hat{c}_j\nonumber\\
&-i\frac{\lambda_R}{a_0^3}\sum_{\braket{ij}}\left(\bm u\cdot\bm d_{ij}\right)\hat{c}_i^{\dagger}\left(\bm s\times\bm d_{ij}\right)_z\hat{c}_j.
\end{align}

For simplicity, we introduce the optical phonons at the Brillouin zone center ($\Gamma$ point), in which the mixture of the longitudinal and transverse optical modes can be circularly polarized~\cite{Zhang2015}.
Here we show the schematic view of the counterclockwise chiral phonon mode at the $\Gamma$ point in Fig.~\ref{fig1}(a), where the sublattices A and B rotate around their equilibrium positions with a phase difference $\pi$, and their displacements are represented by $\bm u_A$ and $\bm u_B$.
The relative displacement vector between the atoms in the A and B sublattices in the modulated Hamiltonian Eqs.~(\ref{dHt}) and (\ref{dHR}) is given by $\bm u=\bm u_B-\bm u_A$. In this case, the Hamiltonian only depends on the relative displacement $\bm u$ between sublattices A and B because there is no phase difference between the same sublattices for the chiral phonon at the $\Gamma$ point.

Then we apply the Fourier transformation $\hat{c}_{\nu\bm r}=\frac{1}{\sqrt{N}}\sum_{\bm k}e^{i\bm k\cdot\bm r}\hat{c}_{\nu\bm k}$, where $\nu=A,B$ labels the two sublattices, and $\hat{c}_{\nu\bm k}=(\hat{c}_{\nu\bm k\uparrow},\hat{c}_{\nu\bm k\downarrow})^T$ denotes the annihilation operator of a Bloch state with a Bloch wavevector $\bm k$. 
Here the primitive vectors $\bm a_{1,2}=(a/2)(\pm1,\sqrt{3})$ with the lattice constant $a$ are depicted in Fig.~\ref{fig1}(a).
Thus, the unmodulated Hamiltonian $\hat{H}_0$ in Eq.~(\ref{H0}) is written as $\hat{H}_0=\sum_{\bm k}\bm v^{\dagger}_{\bm k}\mathcal{H}_0(\bm k)\bm v_{\bm k}$ with the Bloch Hamiltonian $\mathcal{H}_0(\bm k)$ and the basis $\bm v_{\bm k}=(\hat{c}_{A\bm k\uparrow},\hat{c}_{A\bm k\downarrow},\hat{c}_{B\bm k\uparrow},\hat{c}_{B\bm k\downarrow})^T$.
Here we choose the representation of the Dirac matrices as $\Gamma^{1,2,3,4,5}=(\sigma_x\otimes s_0,\sigma_z\otimes s_0,\sigma_y\otimes s_x,\sigma_y\otimes s_y,\sigma_y\otimes s_z)$, where the Pauli matrices $\sigma_{\alpha}$ and $s_{\alpha}$ represent the sublattice index and electron spin, respectively. Their ten commutators are given by $\Gamma^{\alpha\beta}=[\Gamma^{\alpha},\Gamma^{\beta}]/(2i)$. Then the Bloch Hamiltonian is expressed as a linear combination of the Dirac matrices as
\begin{align}
\mathcal{H}_0(\bm k)=~&d_1(\bm k)\Gamma^1+d_{12}(\bm k)\Gamma^{12}+m\Gamma^2+d_{4}(\bm k)\Gamma^{4}\nonumber\\
&+d_{23}(\bm k)\Gamma^{23}+d_{24}(\bm k)\Gamma^{24}+d_{3}(\bm k)\Gamma^{3},
\end{align}
where $d_1=t_0(1+2\cos{x}\cos{y})$, $d_{12}=-2t_0\cos{x}\sin{y}$, $d_{4}=-\sqrt{3}\lambda_R\sin{x}\sin{y}$, $d_{23}=-\lambda_R\cos{x}\sin{y}$, $d_{24}=\sqrt{3}\lambda_R\sin{x}\cos{y}$, and $d_3=\lambda_R(1-\cos{x}\cos{y})$ with $x=k_xa/2$ and $y=\sqrt{3}k_ya$. Figure~\ref{fig1}(c) shows the electronic band structure from the Kane-Mele Hamiltonian in Eq.~(\ref{H0}) without chiral phonons.
Next, the modulated Hamiltonians in Eqs.~(\ref{dHt}) and (\ref{dHR}) can be merged into $\delta\hat{H}(\bm u)=\sum_{\bm k}\bm v^{\dagger}_{\bm k}\delta\mathcal{H}(\bm k,\bm u)\bm v_{\bm k}$ with the Bloch Hamiltonian
\begin{align}
\delta\mathcal{H}(\bm k,\bm u)=~&\delta d_1(\bm k,\bm u)\Gamma^1+\delta d_{12}(\bm k,\bm u)\Gamma^{12}\nonumber\\
&+\delta d_4(\bm k,\bm u)\Gamma^4+\delta d_{23}(\bm k,\bm u)\Gamma^{23}\nonumber\\
&+\delta d_{24}(\bm k,\bm u)\Gamma^{24}+\delta d_3(\bm k,\bm u)\Gamma^3,
\end{align}
where 
\begin{align}
&\delta d_{1}=\frac{t_0}{a_0}\Bigg\{\sqrt{3}u_x\sin{x}\sin{y}+u_y\left(1-\cos{x}\cos{y}\right)\Bigg\}, \nonumber\\
&\delta d_{12}=\frac{t_0}{a_0}\Bigg\{\sqrt{3}u_x\sin{x}\cos{y}+u_y\cos{x}\sin{y}\Bigg\}, \nonumber\\ 
&\delta d_{4}=\frac{\lambda_R}{a_0}\Bigg\{u_x\left(1+\frac{1}{2}\cos{x}\cos{y}\right)+\frac{\sqrt{3}}{2}u_y\sin{x}\sin{y}\Bigg\}, \nonumber\\
&\delta d_{23}=\frac{\lambda_R}{a_0}\Bigg\{\frac{\sqrt{3}}{2}u_x\sin{x}\cos{y}-\frac{3}{2}u_y\cos{x}\sin{y}\Bigg\}, \nonumber\\
&\delta d_{24}=\frac{\lambda_R}{a_0}\Bigg\{\frac{1}{2}u_x\cos{x}\sin{y}-\frac{\sqrt{3}}{2}u_y\sin{x}\cos{y}\Bigg\}, \nonumber\\
&\delta d_{3}=\frac{\lambda_R}{a_0}\Bigg\{-\frac{\sqrt{3}}{2}u_x\sin{x}\sin{y}-\frac{3}{2}u_y\cos{x}\cos{y}\Bigg\}.
\end{align}
Therefore, the total Bloch Hamiltonian with chiral phonons is obtained as 
\begin{align}\label{totH}
\mathcal{H}(\bm k,\bm u)=\mathcal{H}_0(\bm k)+\delta\mathcal{H}(\bm k,\bm u).
\end{align}
When chiral phonons are switched on, the electronic system is periodically driven with time. At the initial transient stage, the system may have energy transfer between phonons and electrons. It will eventually approach a non-equilibrium steady state where the electronic energy modulated by phonons becomes a periodic function of time $t$: $E(t+T)=E(t)$. In Fig.~\ref{fig1}(d), we compare the electronic band structure without phonons (blue dots) and that modulated by phonons (red dots) after taking the time average. There is a small shift of energy which comes from the energy transfer between phonons and electrons at the initial transient stage of phonon-driven electronic dynamics.

\subsection{Berry curvature}
\label{secIIIB}
As we discussed in Sec.~\ref{secII}, when the phonons are represented by one coordinate $\bm u=(u_x,u_y)$, the spin magnetization driven by phonons is given by Eq.~(\ref{mu_z}), expressed in terms of the Berry curvature Eq.~(\ref{BC_n}) in the $u_x$-$u_y$ space.
Meanwhile, since our TB model involves two sublattices with two coordinates $\bm u_A$ and $\bm u_B$, Eq.~(\ref{mu_z}) seems to be unapplicable.
Nevertheless, because the Hamiltonian only depends on the relative position $\bm u\equiv\bm u_B-\bm u_A$ between the atoms in the sublattices A and B, we can still use Eq.~(\ref{mu_z}) by regarding the Berry curvature for the relative displacement $\bm u$.
Thus, the Berry curvature as a function of $\bm u$ for the $n$-th occupied band is given by
\begin{align}\label{BerryCur}
\Omega^{(n)}_{u_xu_y}(\bm k,\bm u)=\sum_{m(\neq n)}\frac{iX^{nm}_{u_x}(\bm k,\bm u)X^{mn}_{u_y}(\bm k,\bm u)+\text{c.c}}{\left[E_{n}(\bm k,\bm u)-E_{m}(\bm k,\bm u)\right]^2},
\end{align}
where $X^{nm}_{u_{\alpha}}(\bm k,\bm u)\equiv\bra{\psi_n(\bm k,\bm u)}\partial_{u_{\alpha}}\mathcal{H}(\bm k,\bm u)\ket{\psi_m(\bm k,\bm u)}$ $(\alpha=x,y)$ with $\ket{\psi_n(\bm k,\bm u)}$ being the $n$-th eigenstate of $\mathcal{H}(\bm k,\bm u)$ in Eq.~(\ref{totH}).

\begin{figure}
\begin{center}
\includegraphics[width=\columnwidth]{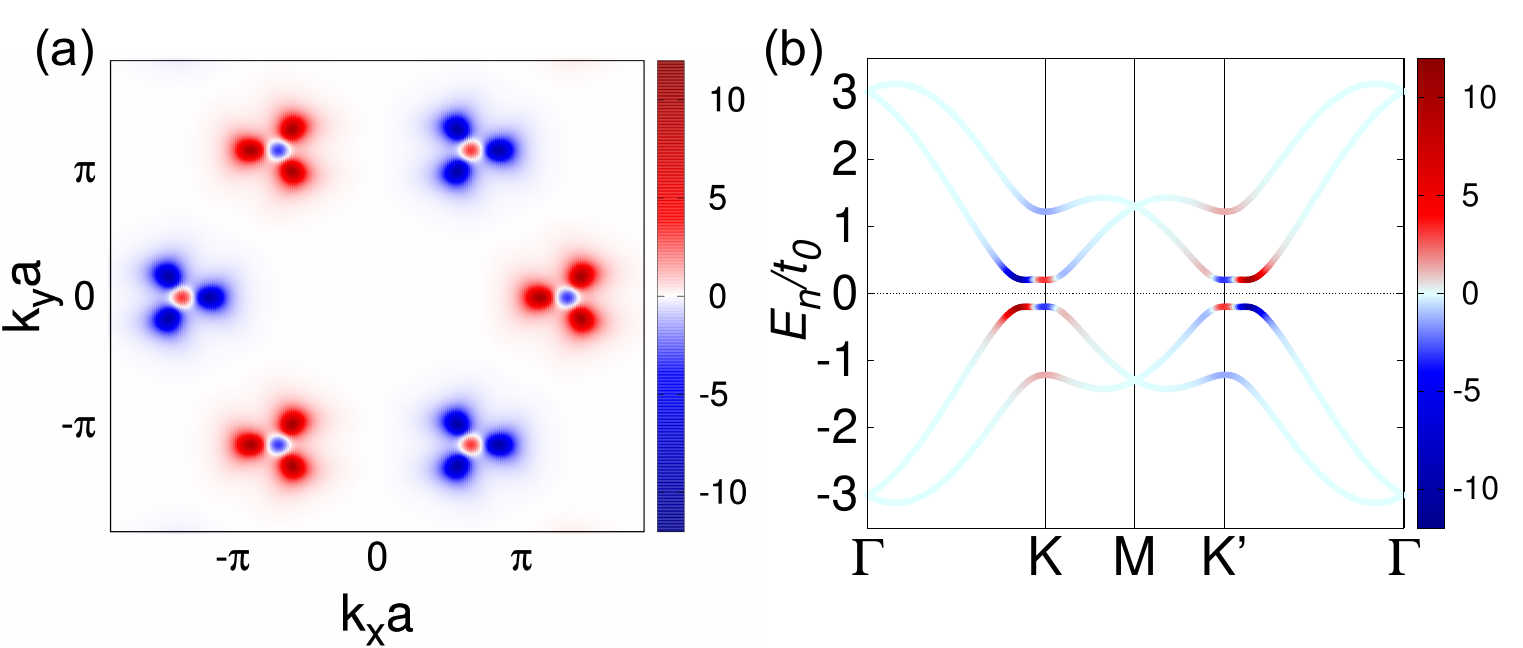}
\caption{(Color online) (a) Distribution of Berry curvature $\Omega_{u_xu_y}^{(n=2)}(k_x,k_y)$ in the $k_x$-$k_y$ space for the second band shown in (b). (b) Electronic band structure along the high-symmetry points with the color denoting the Berry curvature $\Omega_{u_xu_y}^{(n)}$ for each band. The parameters are set to be $m=0.2t_0$, $\lambda_R=0.4t_0$.}
\label{fig2}
\end{center}
\end{figure}

Here we give a simple symmetry analysis on this Berry curvature in Eq.~(\ref{BerryCur}). The total Bloch Hamiltonian $\mathcal{H}(\bm k,\bm u)$ holds threefold rotation symmetry with respect to the $z$ axis because of the hexagonal structure, yielding 
\begin{align}
U\mathcal{H}(\bm k,\bm u)U^{-1}=\mathcal{H}(\bm k',\bm u'),
\end{align}
where $U$ is a unitary matrix given by
\begin{align}
U=
\begin{pmatrix}
1 & 0 \\
0 & e^{-i\bm k\cdot\bm a_2}
\end{pmatrix}\otimes e^{-is_z\frac{\pi}{3}},
\end{align}
and $\hat{C}_3$ represents the threefold rotational matrix, leading to $\bm k'=\hat{C}_3\bm k=(-k_x/2-\sqrt{3}k_y/2,\sqrt{3}k_x/2-k_y/2)$ and $\bm u'=\hat{C}_3\bm u=(-u_x/2-\sqrt{3}u_y/2,\sqrt{3}u_x/2-u_y/2)$.
Then symmetry analysis gives
\begin{align}
X^{nm}_{u_x}(\bm k,\bm u)=&-\frac{1}{2}X^{nm}_{u_x'}(\bm k',\bm u')+\frac{\sqrt{3}}{2}X^{nm}_{u_y'}(\bm k',\bm u'), \\
X^{mn}_{u_y}(\bm k,\bm u)=&-\frac{\sqrt{3}}{2}X^{mn}_{u_x'}(\bm k',\bm u')-\frac{1}{2}X^{mn}_{u_y'}(\bm k',\bm u'),
\end{align}
which leads to
\begin{align}\label{Omega_sym}
\Omega^{(n)}_{u_xu_y}(\bm k,\bm u)=\Omega^{(n)}_{u_x'u_y'}(\bm k',\bm u').
\end{align}
One can also show $\Omega^{(n)}_{u_xu_y}(\bm k',\bm u')|_{\bm u'=0}=\Omega^{(n)}_{u_x'u_y'}(\bm k',\bm u')|_{\bm u'=0}$.
By combining with Eq.~(\ref{Omega_sym}), it follows that $\Omega^{(n)}_{u_xu_y}(\bm k)|_{\bm u=0}=\Omega^{(n)}_{u_xu_y}(\bm k')|_{\bm u=0}$, meaning that the distribution of $\Omega^{(n)}_{u_xu_y}(\bm k)|_{\bm u=0}$ is $C_3$ symmetric in $\bm k$ space.
Figure~\ref{fig2}(a) shows the distribution of $\Omega^{(n=2)}_{u_xu_y}(\bm k)|_{\bm u=0}$ in the Brillouin zone [Fig.~\ref{fig1}(b)] for the upper occupied band shown in Fig.~\ref{fig2}(a) as an example, which clearly shows $C_3$ symmetry. Thus the numerical result agrees with the symmetry analysis $\Omega^{(n)}_{u_xu_y}(\bm k)|_{\bm u=0}=\Omega^{(n)}_{u_xu_y}(\hat{C}_3\bm k)|_{\bm u=0}$.
The distribution of the Berry curvature for each band along the high-symmetry points is depicted in Fig.~\ref{fig2}(b).

\subsection{Spin magnetization}
Next, we calculate the spin magnetization driven by atomic rotations in the TB model by using our formalism based on Eq.~(\ref{mu_z}). In Eq.~(\ref{mu_z}), it is assumed that the phonons are described by a vector $\bm u=(u_x,u_y)$, and as mentioned in Sec.~\ref{secIIIB}, we can apply Eq.~(\ref{mu_z}) by putting $\bm u=\bm u_B-\bm u_A$. It means that the spin magnetization originates from the dynamics of the relative displacement~\cite{Hamada2020,Yao2022,Yao2024}.

\begin{figure}
\begin{center}
\includegraphics[width=\columnwidth]{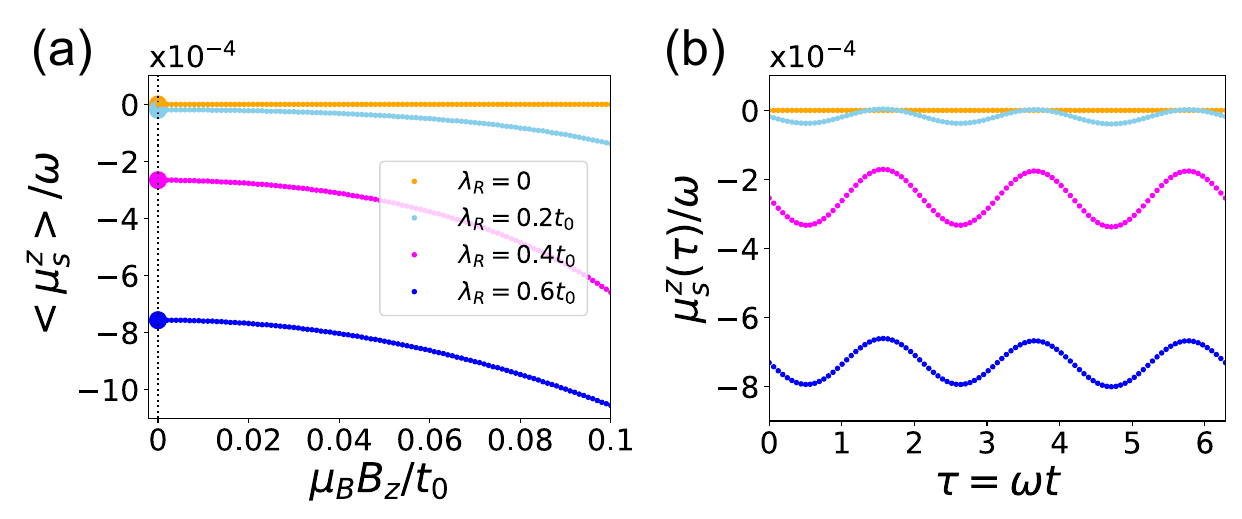}
\caption{(Color online) Results of the calculation on spin magnetization. (a) Phonon-driven spin magnetization $\braket{\mu^z_s}$ calculated by Eq.~(\ref{spinmag_z}) as a function of the Zeeman magnetic field $\bm B=(0,0,B_z)$ with different Rashba terms. Here the values of $\braket{\mu^z_s}$ at $B_z=0$ marked by the large dots represent the spin magnetization driven by chiral phonons without Zeeman magnetic field. (b) Time dependence of the spin magnetization calculated by the method in Ref.~\cite{Hamada2020}. Here we set the relative position $u_r$ to be 4\% of the bond length $a_0$.}
\label{fig3}
\end{center}
\end{figure}	

In the present case, the spin magnetization given by Eq.~(\ref{mu_z}) is proportional to $J_z^{\text{ph}}=\frac{1}{T}\int_0^Tdt(\bm u\times\dot{\bm u})_z$, which is no longer equal to the total phonon angular momentum because $\bm u$ represents the relative displacement $\bm u_B-\bm u_A$. By using Eq.~(\ref{mu_z}), the total spin magnetization for the $z$ component contributed by all the occupied bands is given by
\begin{align}\label{spinmag_z}
\braket{\mu_s^z}=\frac{1}{2}J_z^{\text{ph}}\sum_{n}^{\text{occ}}\int\frac{d\bm k}{(2\pi)^2}\partial_{B_z}\Omega^{(n)}_{u_xu_y}(\bm k)\Big|_{\bm u=0,\bm B=0}.
\end{align}
In this formula, we put $\bm B=0$, but in fact this formula also applies to the cases with $\bm B\neq 0$.
Figure~\ref{fig3}(a) shows the result of phonon-driven spin magnetization $\braket{\mu_z^s}$ as a function of Zeeman magnetic field $\bm B=(0,0,B_z)$ for different values of $\lambda_R$ for the Rashba term. We note that $\braket{\mu_z^s}$ vanishes in the absence of the Rashba SOC.

We note that Eq.~(\ref{spinmag_z}) is obtained as the time average of the time-dependent spin magnetization formulated in Ref.~\cite{Hamada2020}. Next we calculate the time-dependent spin magnetization generated by chiral phonons for our TB model for comparison.
Here, we set the atomic displacements of sublattices A and B corresponding to Fig.~\ref{fig1}(a) to be $\bm u_A=u_A(\cos{\omega t},\sin{\omega t})$, and $\bm u_B=u_B(\cos{\omega t},\sin{\omega t})$, where $\omega$ denotes the phonon frequency. Then their relative displacement is 
\begin{align}\label{re_dis_u}
\bm u=\bm u_B-\bm u_A=u_r(\cos{\tau},\sin{\tau})
\end{align}
where $u_r=u_B-u_A$, and we introduce a dimensionless time $\tau$ by $\tau\equiv\omega t$. The time-dependent spin magnetization is given by Eq.~(\ref{exp_mu}) in Appendix~\ref{Appe_A}.
Meanwhile, $J_z^{\text{ph}}=u_r^2\omega$ is proportional to the phonon frequency $\omega$.
The numerical result of the time-dependent spin magnetization is shown in Fig.~\ref{fig3}(b) with $u_r=0.04a_0$. Indeed, the time averages with different values of $\lambda_R$ for the Rashba term agree well with the results of $\braket{\mu_s^z}$ at $B_z=0$ in Fig.~\ref{fig3}(a).
Thus, our formula Eq.~(\ref{spinmag_z}) gives the time average of the phonon-driven spin magnetization from the Berry phase, formulated in Ref.~\cite{Hamada2020}.
	
\begin{figure}
\begin{center}
\includegraphics[width=\columnwidth]{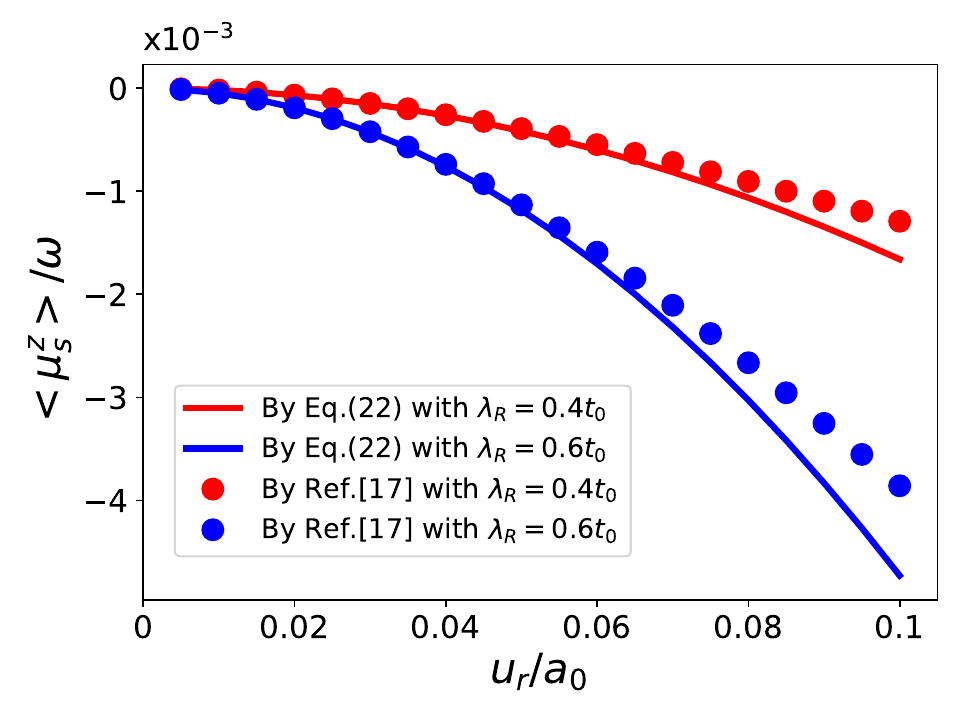}
\caption{(Color online) Dependence of the spin magnetization on the size of displacement. The horizontal axis is the size of displacement $u_r/a_0$, and the vertical axis is the expectation value of spin magnetization $\braket{\mu^z_s}$. Here the colors with red and blue represent the results with the Rashba parameters $\lambda_R=0.4t_0$ and $\lambda_R=0.6t_0$, respectively. The curves and dots represent the results calculated by our formula Eq.~(\ref{mu_z}) and by Ref.~\cite{Hamada2020}, respectively. }
\label{fig4}
\end{center}
\end{figure}	
	
Figure~\ref{fig4} shows the dependence of the expectation value of spin magnetization $\braket{\mu^z_s}$ on the size of displacement $u_r/a_0$. 
The red and blue symbols represent the results with the Rashba parameters $\lambda_R=0.4t_0$ and $\lambda_R=0.6t_0$, respectively. The curves represent the results from our formula Eq.~(\ref{spinmag_z}). For comparison, the red and blue dots show the time averages of the spin magnetization from Ref.~\cite{Hamada2020}.
We notice that they agree well if the displacement is small enough, and they deviate from each other when the displacement $u_r$ compared with the bond length $a_0$ gradually increases.
This is natural since Eq.~(\ref{spinmag_z}) was obtained using an approximation of small $\bm u$. Here we see that the deviation of $\braket{\mu^z_s}$ between our formula Eq.~(\ref{spinmag_z}) and the time average of the spin magnetization in Ref.~\cite{Hamada2020} starts to appear around $u_r/a_0=0.05$. Thus, within our model calculation, our formula Eq.~(\ref{spinmag_z}) is valid when the displacement is less than 5\% of the bond length, which is reasonable for general solids.

\section{Applicability to general cases}
\label{secIV}
In this section, we discuss the applicability of our formalism to general cases.
Since we restricted ourselves to the case with a single atom per unit cell, the phonon displacement is characterized by the coordinate of the single atom in the formalism, where the induced spin magnetization is proportional to the phonon angular momentum of the single atom as we discussed in Sec.~\ref{secII}.
If the unit cell contains more than one atoms, the spin magnetization cannot be directly proportional to the phonon angular momentum.
Nevertheless, in the model calculation, with the $\Gamma$-chiral-phonon mode on the 2D honeycomb lattice shown in Sec.~\ref{secIII}, our formula Eq.~(\ref{mu_z}) is still applicable because the modulated Hamiltonian depends only on the relative displacement between two sublattices. 
It is noted that $J_z^{\text{ph}}$ is no longer the total phonon angular momentum.
In this case, the Berry curvature $\Omega^{(n)}_{u_xu_y}$ for the relative displacement $\bm u=(u_x,u_y)$ determines the spin magnetization.

In addition, chiral phonons can appear at high-symmetry points on other typical lattices, such as the 2D kagome lattice~\cite{Chen2019} and the lattice with $C_4$ symmetry~\cite{Wang2022}. 
In the case of the 2D kagome lattice, if we consider the chiral phonons at the $\Gamma$ point~\cite{Chen2019,Yao2024}, since the unit cell contains three sublattices A, B, and C, the modulated Hamiltonian depends on two relative displacement $\bm u_{AB}$ and $\bm u_{BC}$.
Thus, the cross terms from different relative displacement in $J_z^{\text{ph}}$ contributes to the spin magnetization, and Eq.~(\ref{mu_z}) should be revised as
\begin{align}\label{generalfor}
\braket{\mu^{\alpha}_{s,n}}=\frac{1}{2}\sum_{a,b=1}^{N-1}J^{\text{ph}}_{z,ab}\int\frac{d\bm k}{(2\pi)^d}\partial_{B_{\alpha}}\Omega^{(n)}_{u_{ax}u_{by}}\Big|_{\bm u_{a,b}=0,\bm B=0},
\end{align}
where $u_{ai}$ ($a=1,\cdots,N-1$ and $i=x,y,z$) stands for the relative displacement between different sublattices, $a,b$ denotes the degrees of freedom of the relative displacements, $J^{\text{ph}}_{z,ab}=\frac{1}{T}\int_0^Tdt(\bm u_a\times\dot{\bm u_b})_z$ represents the contribution from these cross terms, and $N$ is the number of atoms per unit cell. Obviously, at the $\Gamma$ point, $N=2$ for the chiral phonon on the 2D honeycomb lattice as we shown in Sec.~\ref{secIII}, and $N=3$ for the chiral phonon on the 2D kagome lattice.
On the other hand, if we consider the chiral phonons at valley ($K$ or $K'$) points, in which the same atoms located at the different unit cells rotate with different phases~\cite{Zhang2015,Chen2019}, the valley phonons can always be folded onto the $\Gamma$ point by appropriately enlarging the unit cell~\cite{Ren2015}.
Therefore, at the valley points, $N=6$ for the chiral phonon on the 2D honeycomb lattice, and $N=9$ for the chiral phonon on the 2D kagome lattice.
Here $\Omega^{(n)}_{u_{ax}u_{by}}\equiv\partial_{u_{ax}}A^{(n)}_{u_{by}}-\partial_{u_{by}}A^{(n)}_{u_{ax}}$ is the Berry curvature corresponding to these cross terms, and it is determined by the relative positions of the different atoms on the lattice.
Similarly, phonons with wavevectors equal to a fraction of high-symmetry wavevectors can be treated in the same manner. Meanwhile, our theory cannot be applied to phonons with arbitrary wavevectors in general, and it is left as a future work.

For real materials, we can calculate $\Omega^{(n)}_{u_{ax}u_{by}}$ for the $n$-th occupied electronic band by $ab$ $initio$ calculations. Then we can evaluate how much spin magnetization can be driven by chiral phonons in real materials by using Eq.~(\ref{generalfor}).
On the other hand, some recent experiments already show generation of spin magnetization by chiral phonons. For example, a spin Seebeck effect due to chiral phonons is observed~\cite{Kim2023}, and a phonon angular momentum in a chiral quartz crystal is measured via the inverse spin Hall effect~\cite{Ohe2024}. Both measurements are attributed to generation of spin magnetization by chiral phonons, which is the main topic of the present paper. Meanwhile, in these measurements, the chiral phonons are generated by thermal gradient, and this generation method by thermal gradient (phonon thermal Edelstein effect)~\cite{Hamada2018} will induce chiral phonons with various wavevectors and frequencies. Since our theory is limited to some special values of wavevectors, we cannot evaluate the magnitude of the signal in these measurements, and it is left as a future work.

\section{Conclusion and discussion}
\label{secV}	
In this paper, we construct a general theory of spin magnetization driven by chiral phonons under an adiabatic process, in which atoms rotate around their equilibrium positions with a low phonon frequency. The spin magnetization originates from the modulated electronic states with SOC by atomic rotations. 
Under the adiabatic approximation, the time-dependent spin magnetization can be calculated by the Berry-phase method~\cite{Hamada2020}. Here we focus on its time average, which is evaluated by assuming that the phonon displacement is small. As a result, the time average of the spin magnetization is concisely related to the phonon angular momentum, and encoded in the Berry curvature defined in the phonon-displacement space.
Our results well agree with that obtained from the time-dependent description~\cite{Hamada2020} with a realistic size of atomic displacements. 
The merit of this methodology is that we can still simply deal with the interaction between the chiral phonons and electrons by a single-particle picture for electrons. The effect of chiral phonons is naturally reflected in electronic states due to the geometric effect.

We note that another theory on a chiral phonon-spin conversion~\cite{Funato2024} has been proposed, based on the spin-rotation coupling (SRC)~\cite{Matsuo2011,MatsuoPRB2011,Matsuo2013}. This SRC has been derived from the Dirac equation seen from the rotating frame~\cite{Matsuo2011}, and it says that there is a direct coupling between rotational motion and spin magnetization. Nonetheless, it is not obvious whether the SRC applies to a coupling between chiral phonons and electron spins, because its microscopic origin still needs further study. 
Furthermore, we note that even when chiral phonons are present, if seen from the particular point where an electron resides, nuclei are moving with different angular velocities. Therefore, we suspect that the SRC in solids is an effective theory, which is valid when the electrons are strongly bound to nuclei. Remarkably, this SRC does not need the SOC to generate electron spins. This point is in distinct contrast to our paper, which requires SOC, and these scenarios can be distinguished in experiments.

In our formalism, because the spin magnetization driven by chiral phonons is proportional to the time average of the phonon angular momentum, only the contribution from the rotational modes becomes nonzero. Moreover, our formalism reflects the chiral nature of phonons, corresponding to the spin magnetization with opposite signs driven by the clockwise and counterclockwise rotational modes. Meanwhile, the spin magnetization vanishes in simple vibration modes. Our formalism is applicable to general cases and is convenient for $ab$ $initio$ calculations.

It should be noted that in nonmagnetic systems, the phonon angular momenta at $\bm k$ and $-\bm k$ are opposite due to the time-reversal symmetry, and therefore if these two modes are equally populated, spin magnetization will not be induced.
By phonon pumping with terahertz pulses~\cite{Nova2017}, which leads to an asymmetric population of chiral phonons between clockwise and counterclockwise modes, one can obtain a nonzero net spin magnetization.
Furthermore, by using the same formalism, one can formulate the chiral phonon-induced orbital magnetization~\cite{Trifunovic2019,Xiao2021}, electric current~\cite{Yao2022}, and magnon excitations~\cite{Yao2024} proposed in the previous studies. The time averages of these physical quantities driven by low-frequency chiral phonons yield a similar formulation.

\begin{acknowledgments}
This work was supported by Japan Society for the Promotion of Science (JSPS) KAKENHI Grants No.~JP20H04633, No.~JP22H00108, No.~JP24H02231, and also by MEXT Initiative to Establish Next-generation Novel Integrated Circuits Centers (X-NICS) Grant No.~JPJ011438.
D.Y. was also supported by JSPS KAKENHI Grant No.~JP23KJ0926.
\end{acknowledgments}

\begin{appendix}

\section{General formula of the geometric spin magnetization}
\label{Appe_A}
In this Appendix, we show the derivation of Eq.~(\ref{mu_sn}) in the main text.
The Hamiltonian describing a coupling between the electron spins and the Zeeman magnetic field $\bm B$ is given by
\begin{align}\label{ZeemanH}
\hat{H}_B=-\hat{\bm \mu}_s\cdot\bm B,
\end{align}
where $\hat{\bm \mu}_s=-\mu_B\hat{\bm s}$ represents the spin magnetization with the Bohr magneton $\mu_B$ and the Pauli matrices $\hat{s}_{\alpha}$ $(\alpha=x,y,z)$.
In the absence of the Zeeman magnetic field, the spin magnetization for the $\alpha$ component is given by
\begin{align}
\hat{\mu}^{\alpha}_s=-\partial_{B_{\alpha}}\hat{H}|_{\bm B=0},
\end{align}
where $\hat{H}$ is the total Hamiltonian. As was derived in Ref.~\cite{Hamada2020}, by using the adiabatic approximation, the geometric term of the expectation value of $\hat{\mu}^{\alpha}_s$ for the $n$-th electronic band to the first order is given by
\begin{align}\label{exp_mu}
\mu^{\alpha}_{s,n}(t)=\int\frac{d\bm k}{(2\pi)^d}\sum_{m(\neq n)}\left[\frac{-\hat{\mu}^{\alpha}_{s,nm}(\bm k,t)A_{mn}(\bm k,t)}{E_{n}(\bm k,t)-E_{m}(\bm k,t)}+\text{c.c}\right].
\end{align}
Here 
\begin{align}\label{munm}
&\hat{\mu}^{\alpha}_{s,nm}(\bm k,t)=\bra{\psi_{n}(\bm k,t)}\left(-\partial_{B_{\alpha}}\hat{H}|_{\bm B=0}\right)\ket{\psi_{m}(\bm k,t)} \nonumber\\
=&\left[E_{n}(\bm k,t)-E_{m}(\bm k,t)\right]\braket{\psi_{n}(\bm k,t)|\partial_{B_{\alpha}}\psi_{m}(\bm k,t)}\Big|_{\bm B=0}
\end{align}
represents the instantaneous matrix element of $\hat{\mu}^{\alpha}_{s}$ at time $t$, and $A_{nm}(\bm k,t)=i\bra{\psi_{n}(\bm k,t)}\partial_t\ket{\psi_{m}(\bm k,t)}$ is the instantaneous Berry connection, where $\ket{\psi_{n}(\bm k,t)}$ is the eigenstate of the $n$-th band for the wavevector $\bm k$ at time $t$.
In Ref.~\cite{Hamada2020}, the time dependence of the spin magnetization is calculated.

In the case of the atomic rotation with the displacement $\bm u=(u_{x},u_{y})$, the Berry connection can be replaced by
\begin{align}\label{Amn}
A_{nm}(\bm k,\bm u)=i\braket{\psi_{n}(\bm k,\bm u)|\partial_{\bm u}\psi_{m}(\bm k,\bm u)}\cdot\dot{\bm u}.
\end{align}
By substituting Eqs.~(\ref{munm}) and (\ref{Amn}) into Eq.~(\ref{exp_mu}), the spin magnetization driven by the atomic rotation can be expressed as
\begin{align}
\mu^{\alpha}_{s,n}=&\sum_{\delta=x,y}\dot{u}_{\delta}\int\frac{d\bm k}{(2\pi)^d}\Big[i\braket{\partial_{B_{\alpha}}\psi_n|\partial_{u_{\delta}}\psi_n}+\text{c.c}\Big]\Big|_{\bm B=0} \nonumber\\
=&\sum_{\delta=x,y}\dot{u}_{\delta}\int\frac{d\bm k}{(2\pi)^d}\Omega^{(n)}_{B_{\alpha}u_{\delta}}\Big|_{\bm B=0},
\end{align}
where $\Omega^{(n)}_{B_{\alpha}u_{\delta}}=\partial_{B_{\alpha}}A^{(n)}_{u_{\delta}}-\partial_{u_{\delta}}A^{(n)}_{B_{\alpha}}$ is the Berry curvature with $A^{(n)}_{\rho}=i\braket{\psi_n|\partial_{\rho}\psi_n}$ $(\rho=u_{\delta},B_{\alpha})$ being the Berry connection in a general form~\cite{Xiao2009}.

\section{Calculation of the Zeeman magnetic field derivative of the Berry curvature}
\label{Appe_B}
In this Appendix, we derive a formula for the derivative of the Berry curvature $\partial_{B_{\alpha}}\Omega_{u_xu_y}^{(n)}$ in terms of the Zeeman magnetic field in Eq.~(\ref{mu_z}) obtained in the main text, where the Berry curvature $\Omega_{u_xu_y}^{(n)}$ is given by Eq.~(\ref{BerryCur}). The derivative reads
\begin{align}\label{partB}
\partial_{B_{\alpha}}&\Omega_{u_xu_y}^{(n)}=\sum_{m}^{\text{unocc}}\Bigg[i\frac{(\partial_{B_{\alpha}}X_{u_x}^{nm})X_{u_y}^{mn}+X_{u_x}^{nm}(\partial_{B_{\alpha}}X_{u_y}^{mn})}{(E_n-E_m)^2} \nonumber\\
&~~-2i\frac{X_{u_x}^{nm}X_{u_y}^{mn}(\partial_{B_{\alpha}}E_n-\partial_{B_{\alpha}}E_m)}{(E_n-E_m)^3}+\text{c.c}\Bigg],
\end{align}
where
\begin{align}
\partial_{B_{\alpha}}X_{u_{\beta}}^{nm}=&\bra{\partial_{B_{\alpha}}\psi_n}\partial_{u_{\beta}}\hat{H}\ket{\psi_m}+\bra{\psi_n}\partial_{u_{\beta}}\hat{H}\ket{\partial_{B_{\alpha}}\psi_m} \nonumber\\
&+\bra{\psi_n}\partial_{B_{\alpha}}\partial_{u_{\beta}}\hat{H}\ket{\psi_m} \nonumber\\
=&\sum_{l(\neq n)}\frac{\bra{\psi_n}\partial_{B_{\alpha}}\hat{H}\ket{\psi_l}\bra{\psi_l}\partial_{u_{\beta}}\hat{H}\ket{\psi_m}}{E_n-E_l} \nonumber\\
&+\sum_{l(\neq m)}\frac{\bra{\psi_n}\partial_{u_{\beta}}\hat{H}\ket{\psi_l}\bra{\psi_l}\partial_{B_{\alpha}}\hat{H}\ket{\psi_m}}{E_m-E_l} \nonumber\\
=&~\mu_B\sum_{l(\neq n)}\frac{\hat{s}^{nl}_{\alpha}X^{lm}_{u_{\beta}}}{E_n-E_l}+\mu_B\sum_{l(\neq m)}\frac{X^{nl}_{u_{\beta}}\hat{s}^{lm}_{\alpha}}{E_m-E_l}, 
\end{align}
with $\beta=x,y$, and
\begin{align}
\partial_{B_{\alpha}}E_n=&\bra{\psi_n}\partial_{B_{\alpha}}\hat{H}\ket{\psi_n}=\mu_B\hat{s}^{nn}_{\alpha}.
\end{align}
Here we have embedded the identity operator $\mathbb{I}=\sum_{l}\ket{\psi_l}\bra{\psi_l}$, and made use of the Sternheimer identity 
\begin{align}
(E_n-E_l)\braket{\psi_l|\partial_{B_{\alpha}}\psi_n}=\bra{\psi_l}\partial_{B_{\alpha}}\hat{H}\ket{\psi_n},
\end{align}
with $l\neq n$, which is obtained by differentiating $\hat{H}\ket{\psi_n}=E_n\ket{\psi_n}$. Meanwhile, we have replaced the Zeeman magnetic field derivative of the Hamiltonian $\hat{H}$ by $\partial_{B_{\alpha}}\hat{H}=\mu_B\hat{s}_{\alpha}$ from Eq.~(\ref{ZeemanH}).
Therefore, Eq.~(\ref{partB}) is given by
\begin{align}
\partial_{B_{\alpha}}&\Omega^{(n)}_{u_xu_y}=i\mu_B\sum_m^{\text{unocc}}\sum_{l(\neq n)}\frac{\hat{s}_{\alpha}^{nl}X_{u_x}^{lm}X_{u_y}^{mn}+X_{u_x}^{nm}X_{u_y}^{ml}\hat{s}_{\alpha}^{ln}}{(E_n-E_m)^2(E_n-E_l)}\nonumber\\
&+i\mu_B\sum_m^{\text{unocc}}\sum_{l(\neq m)}\frac{X_{u_x}^{nl}\hat{s}_{\alpha}^{lm}X_{u_y}^{mn}+X_{u_x}^{nm}\hat{s}_{\alpha}^{ml}X_{u_y}^{ln}}{(E_n-E_m)^2(E_m-E_l)}\nonumber\\
&-2i\mu_B\sum_{m}^{\text{unocc}}\frac{X_{u_x}^{nm}X_{u_y}^{mn}}{(E_n-E_m)^3}\left(\hat{s}_{\alpha}^{nn}-\hat{s}_{\alpha}^{mm}\right)+\text{c.c}.
\end{align}
By taking a summation for $n$ over the occupied bands, the summations over $l$ can further be rewritten as
\begin{align}
&\sum_n^{\text{occ}}\partial_{B_{\alpha}}\Omega^{(n)}_{u_xu_y}\nonumber\\
=&~i\mu_B\sum_n^{\text{occ}}\sum_m^{\text{unocc}}\sum_{l}^{\text{unocc}}\frac{\hat{s}_{\alpha}^{nl}X_{u_x}^{lm}X_{u_y}^{mn}+X_{u_x}^{nm}X_{u_y}^{ml}\hat{s}_{\alpha}^{ln}}{(E_n-E_m)^2(E_n-E_l)}\nonumber\\
+i&\mu_B\sum_n^{\text{occ}}\sum_m^{\text{unocc}}\sum_{l}^{\text{occ}}\frac{X_{u_x}^{nl}\hat{s}_{\alpha}^{lm}X_{u_y}^{mn}+X_{u_x}^{nm}\hat{s}_{\alpha}^{ml}X_{u_y}^{ln}}{(E_n-E_m)^2(E_m-E_l)}\nonumber\\
-2i&\mu_B\sum_n^{\text{occ}}\sum_{m}^{\text{unocc}}\frac{X_{u_x}^{nm}X_{u_y}^{mn}}{(E_n-E_m)^3}\left(\hat{s}_{\alpha}^{nn}-\hat{s}_{\alpha}^{mm}\right)+\text{c.c}.
\end{align}
This formula allows us to calculate Eq.~(\ref{mu_z}) without taking the $B_{\alpha}$ derivative.

\end{appendix}


%

\end{document}